\documentclass[twocolumn]{revtex4}
\usepackage{graphicx}

\def\duzomniejsze{<\kern-.7mm<}
\def\duzowieksze{>\kern-.7mm>}

\def\textbf#1{{\bf #1}}
\def\beq{\begin{equation}}
\def\eeq{\end{equation}}
\def\be{\begin{equation}}
\def\ee{\end{equation}}
\def\ben{\begin{eqnarray}}
\def\een{\end{eqnarray}}
\def\beqa{\begin{eqnarray}}
\def\eeqa{\end{eqnarray}}
\def\eea{\end{array}}
\def\bea{\begin{array}}
\newcommand{\bei}{\begin{itemize}}
\newcommand{\eei}{\end{itemize}}
\newcommand{\bee}{\begin{enumerate}}
\newcommand{\eee}{\end{enumerate}}

\def\>{\rangle}
\def\<{\langle}

\begin{document}

\title{Quantum control of entanglement and information of two solid state qubits: remote control of dephasing}

\begin{abstract}
We investigate the scheme for controlling information
characterized by Von-Neumann entropy and the stationary state
entanglement characterized by concurrence of two solid state
qubits in the collective dephasing channel. It is shown that the
local maximal value of the stationary state concurrence always
corresponds to the local minimal value of information. We also
propose a scheme for remotely controlling the entanglement of two
solid state qubits against the collective dephasing. This idea may
open a door to remotely suppress the detrimental effects of
decoherence.

\end{abstract}
\author{Shang-Bin Li}\email{sbli@zju.edu.cn},\author{Jing-Bo Xu}

\affiliation{Chinese Center of Advanced Science and Technology
(World Laboratory), P.O.Box 8730, Beijing, People's Republic of
China;} \affiliation{Zhejiang Institute of Modern Physics and
Department of Physics, Zhejiang University, Hangzhou 310027,
People's Republic of China}

\maketitle

\section * {I. INTRODUCTION}

Quantum entanglement plays an important role in quantum
information processes, which can exhibit the nature of a nonlocal
correlation between quantum systems that have no classical
interpretation \cite{Nielsen2000}. Recently, it has been
recognized that entanglement can be used as an important resource
for quantum teleportation or quantum computation. Ordinarily,
entanglement can be destroyed by the interaction between quantum
systems of interest and its surrounding environments. Certain kind
of the interaction between the physical system and environments or
measuring apparatus can lead to the collective dephasing, which
occurs in the physical systems such as trapped ions, quantum dots.
Collective dephasing allows the existence of the so-called
decoherence-free subspace \cite{Zanardi2001}. Several strategies
have been proposed to suppress the detrimental effects of
decoherence, while at the same time allowing for robust
manipulation of the quantum information
\cite{Shor1995,Steane1996,Gottesman1996,Ekert1996,Calderband1997,Plenio1997,Vitali1997,Duan1997,Bacon2000,Viola1998,Poyatos1996},
for example, the quantum error correction schemes
\cite{Shor1995,Steane1996,Gottesman1996,Ekert1996,Calderband1997,Plenio1997},
feedback implementations \cite{Vitali1997}, quantum error avoiding
approach \cite{Duan1997,Bacon2000}, dynamical decoupling
techniques \cite{Viola1998}, engineering of pointer state methods
\cite{Poyatos1996}. Here, we propose a scheme for remotely
controlling the dephasing of two qubits. It is shown that, via a
preexistent multi-partite entanglement and the quantum erasing
process, we can remotely control the entanglement of two qubits in
the collective dephasing channel.

Recently, the quantum information processes in the presence of the
collective dephasing have intrigued much attention
\cite{Khodjasteh2002,Yu2002,Yu2003,Hill2004}. Khodjasteh and Lidar
have investigated the universal fault-tolerant quantum computation
in the presence of spontaneous emission and collective dephasing
\cite{Khodjasteh2002}. Hill and Goan have studied the effect of
dephasing on proposed quantum gates for the solid-state Kane
quantum computing architecture \cite{Hill2004}.It has been shown
that some kinds of entangled states of two qubits are very fragile
in the presence of collective dephasing and the others are very
robust \cite{Yu2002}. So, it is very important to protect the
fragile entangled states from completely losing their
entanglement. Here, we comparatively investigate the information
characterized by Von-Neumann entropy and the stationary state
entanglement quantified by concurrence of two locally driven
qubits in the presence of collective dephasing. We show that one
can transform the fragile entangled states into the stationary
entangled states under the collective dephasing by making use of a
finite-time external driving field. The local maximal value of the
stationary state concurrence corresponds to the local minimal
value of information.

The disentanglement of entangled states of qubits is also a very
important issue for quantum information processes, such as the
solid state quantum computation. For example, in quantum
registers, some kinds of undesirable entanglement between the
qubits can lead to the decoherence of the qubit \cite{Reina2002}.
Yu and Eberly have found that the time for decay of the qubit
entanglement can be significantly shorter than the time for local
dephasing of the individual qubits \cite{Yu2002,Yu2003}. The
collective dephasing can be described by the master equation \be
\frac{\partial\hat{\rho}}{\partial{t}}=\frac{\gamma}{2}(2\hat{J}_{z}\hat{\rho}\hat{J}_{z}-\hat{J}^2_{z}\hat{\rho}-\hat{\rho}\hat{J}^2_{z}),
\ee where $\gamma$ is the decay rate. $\hat{J}_{z}$ are the
collective spin operator defined by \be
\hat{J}_{z}=\sum^{2}_{i=1}\hat{\sigma}^{(i)}_{z}/2,\ee where
$\hat{\sigma}_z$ for each qubit is defined by
$\hat{\sigma}_{z}=|1\rangle\langle{1}|-|0\rangle\langle{0}|$.
Previous studies have shown that two of the four Bell states
$|\Psi^{\pm}\rangle\equiv\frac{\sqrt{2}}{2}(|11\rangle\pm|00\rangle)$
are fragile states in the collective dephasing channel, while the
others Bell states
$|\Phi^{\pm}\rangle\equiv\frac{\sqrt{2}}{2}(|10\rangle\pm|01\rangle)$
are robust entangled states.

This paper is organized as follows: In section II, we study the
system in which one of two solid state qubits in a collective
dephasing environment is driven by a finite-time external field.
We compare the information characterized by Von-Neumann entropy
and the entanglement characterized by concurrence of the
stationary state and find that the local maximal value of the
stationary state concurrence always corresponds to the local
minimal value of information. In section III, we investigate how
to remotely control the dephasing of two qubits by making use of a
preexistent multi-partite entanglement and the quantum erasing
process. In section IV, there are some conclusions.

\section * {II. INFORMATION AND ENTANGLEMENT IN THE STATIONARY STATE OF TWO SOLID STATE QUBITS}

Here, we investigate the model in which two solid state qubits are
exposed in a collective dephasing channel and one of two qubits is
simultaneously driven by a finite-time external field. The
dynamics of two qubits can be described by the following master
equation \be
\frac{\partial\hat{\rho}}{\partial{t}}=-\frac{i}{2}[\Omega_1(t)\hat{\sigma}^{(1)}_x,\hat{\rho}]+\frac{\gamma}{2}(2\hat{J}_{z}\hat{\rho}\hat{J}_{z}-\hat{J}^2_{z}\hat{\rho}-\hat{\rho}\hat{J}^2_{z}),
\ee where $\Omega_{1}(t)=\Omega_1\Theta(T-t)$ is the intensity of
the time-dependent external driving field acted on the qubit 1,
and $\Theta(x)$ is the unit step function and equals one for
$x\geq0$ and equals zero for $x<0$.
$\sigma^{(1)}_x\equiv|1\rangle_1\langle0|+|0\rangle_1\langle1|$ is
one of the Pauli matrices. In the following calculations, we will
show that the action time $T$ of the external driving field plays
a special role in the stationary state entanglement of two qubits.
After numerically solving the master equation (3), we can obtain
the stationary state density matrix of two qubits. Without loss of
generality, the stationary state density matrix has the form \beqa
\rho_s&=&a(\frac{\Omega_1}{\gamma},\gamma{T})|11\rangle\langle11|+b(\frac{\Omega_1}{\gamma},\gamma{T})|10\rangle\langle10|\nonumber\\
&&+c(\frac{\Omega_1}{\gamma},\gamma{T})|01\rangle\langle01|+d(\frac{\Omega_1}{\gamma},\gamma{T})|00\rangle\langle00|\nonumber\\
&&+f(\frac{\Omega_1}{\gamma},\gamma{T})|10\rangle\langle01|+f^{\ast}(\frac{\Omega_1}{\gamma},\gamma{T})|01\rangle\langle10|.
\eeqa In order to quantify the degree of entanglement, we adopt
the concurrence $C$ defined by Wooters \cite{Woo1998}. The
concurrence varies from $C=0$ for an unentangled state to $C=1$
for a maximally entangled state. For two qubits, in the "Standard"
eigenbasis: $|1\rangle\equiv|11\rangle$,
$|2\rangle\equiv|10\rangle$, $|3\rangle\equiv|01\rangle$,
$|4\rangle\equiv|00\rangle$, the concurrence may be calculated
explicitly from the following: \be
C=\max\{\lambda_1-\lambda_2-\lambda_3-\lambda_4,0\}, \ee where the
$\lambda_{i}$($i=1,2,3,4$) are the square roots of the eigenvalues
\textit{in decreasing order of magnitude} of the "spin-flipped"
density matrix operator
$R=\rho_s(\sigma^{y}\otimes\sigma^{y})\rho^{\ast}_s(\sigma^{y}\otimes\sigma^{y})$,
where the asterisk indicates complex conjugation. The concurrence
related to the density matrix $\rho_s$ can be written as \be
C_s=2\max[0,|f(\frac{\Omega_1}{\gamma},\gamma{T})|-\sqrt{a(\frac{\Omega_1}{\gamma},\gamma{T})d(\frac{\Omega_1}{\gamma},\gamma{T})}].
\ee The information quantified by the Von-Neumann entropy of the
stationary state of two qubits is defined by
$S=-{\mathrm{Tr}}(\rho_s\log_2\rho_s)$. We can easily obtain the
analytical expression of the information $S$ as follows, \beqa
S&=&-a\log_2a-d\log_2d\nonumber\\
&&-\beta_{+}\log_2\beta_{+}-\beta_{-}\log_2\beta_{-},\eeqa where
\be \beta_{\pm}=\frac{b+c\pm\sqrt{(b-c)^2+4|f|^2}}{2}. \ee
Firstly, we consider the case in which two qubits are initially in
the Bell state $|\Phi^{-}\rangle$, i.e., the Bell singlet state.
\begin{figure}
\centerline{\includegraphics[width=2.5in]{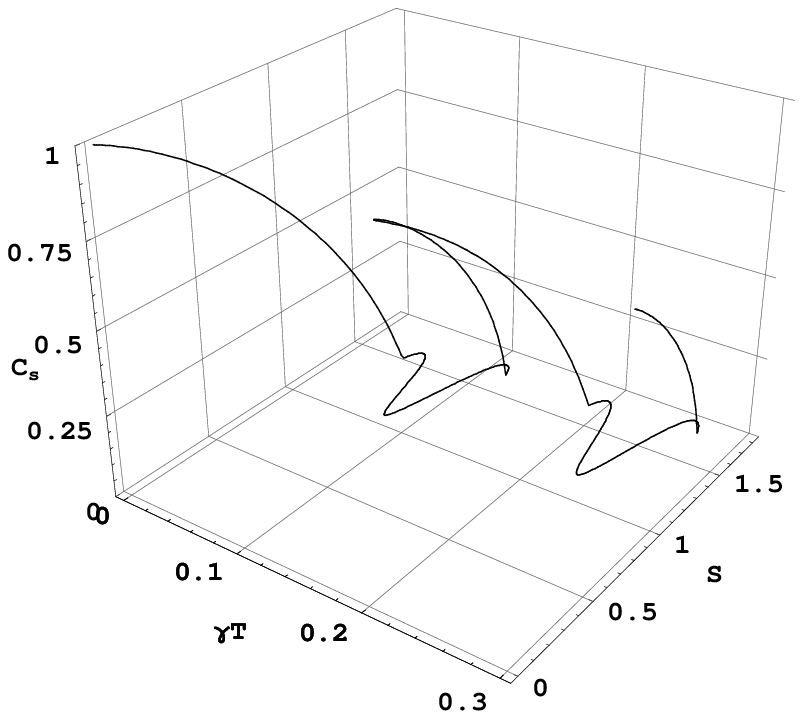}}
\caption{The stationary state concurrence $C_s$ and the
information $S$ of the stationary state are plotted as the
function of the parameter $\gamma{T}$ with
$\frac{\Omega_1}{\gamma}=41.25$. In this case, two qubits are
initially in the Bell state $|\Phi^{-}\rangle$. }
\end{figure}

In Fig.1, the stationary state concurrence and the information of
the stationary state are plotted as the function of the parameter
$\gamma{T}$. It is shown that the values of the scaled action time
$\gamma{T}$ which locally maximizes the stationary state
entanglement always locally minimizes the information. Both the
information of the stationary state and the stationary state
concurrence oscillate with $\gamma{T}$, which imply one can
control both the entanglement and information of the stationary
state of two qubits in the collective dephasing environment by
adjusting the scaled action time $\gamma{T}$ of the locally
driving field. Then, we consider the case in which two qubits are
initially in the standard Werner state. The standard two-qubit
Werner state was defined by \cite{Werner1989} \be
\rho_W=r|\Phi^{-}\rangle\langle\Phi^{-}|+\frac{1-r}{4}I\otimes{I},
\ee where $r\in[0,1]$, and $I$ is the identity operator of a
single qubit. The concurrence of the standard Werner state
decreases with its information. In Fig.2, we display the
stationary state concurrence and the information of the stationary
state as the function of the parameter $\gamma{T}$. Two qubits are
initially in
$0.8|\Phi^{-}\rangle\langle\Phi^{-}|+0.05I\otimes{I}$. We can see
that the initial mixedness changes the stationary state
concurrence and information even though $\Omega_1/\gamma$ and
$\gamma{T}$ are both fixed. While the correspondence between the
local maximal value of the concurrence and the local minimal value
of the information is also valid.

\begin{figure}
\centerline{\includegraphics[width=2.5in]{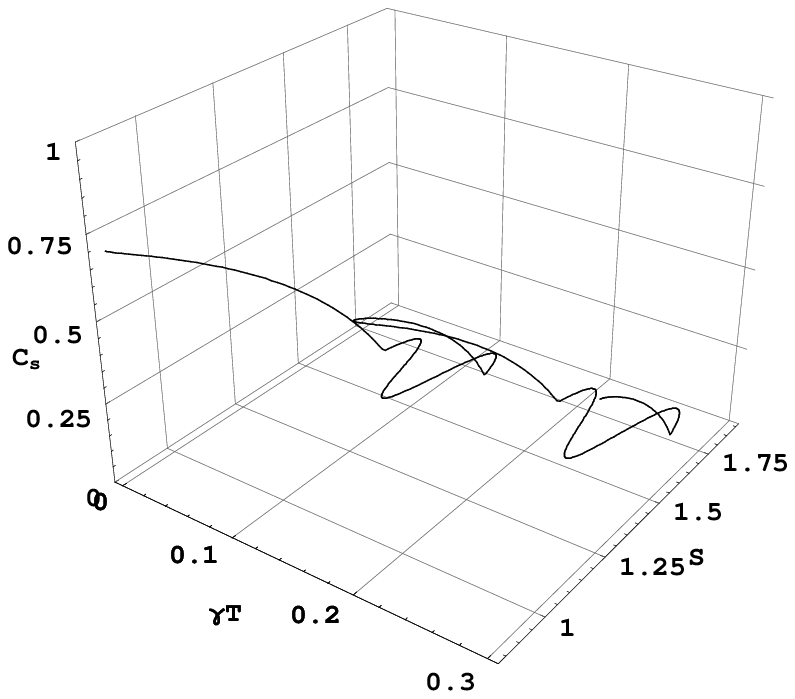}}
\caption{The stationary state concurrence $C_s$ and the
information $S$ of the stationary state are plotted as the
function of the parameter $\gamma{T}$ with
$\frac{\Omega_1}{\gamma}=41.25$. In this case, two qubits are
initially in the standard Werner state $\rho_W$ with $r=0.8$. }
\end{figure}

\begin{figure}
\centerline{\includegraphics[width=2.5in]{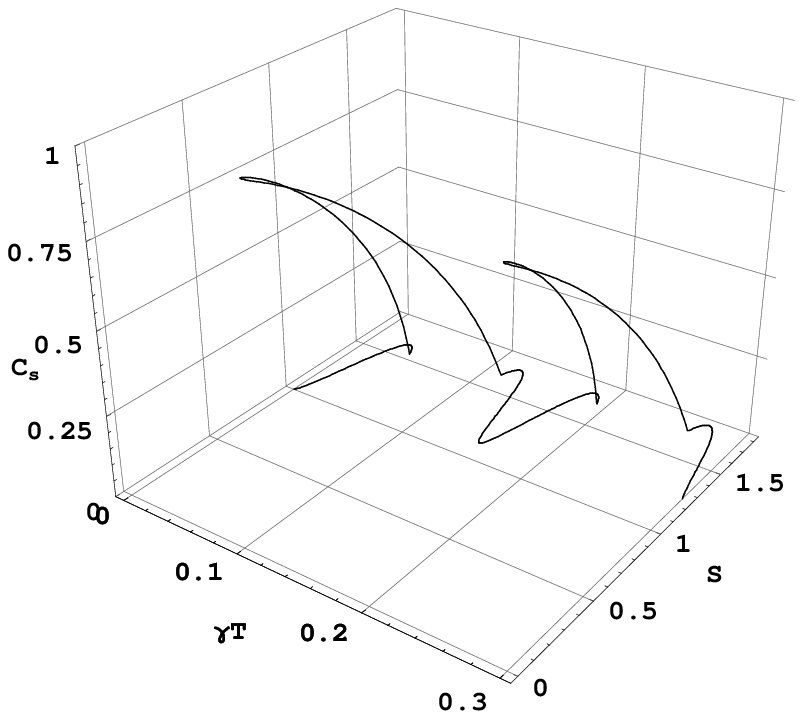}}
\caption{The stationary state concurrence $C_s$ and the
information $S$ of the stationary state are plotted as the
function of the parameter $\gamma{T}$ with
$\frac{\Omega_1}{\gamma}=41.25$. In this case, two qubits are
initially in the Bell state $|\Psi^{+}\rangle$.}
\end{figure}

At the end of this section, we consider the case in which two
qubits are initially in the fragile entangled state
$|\Psi^{+}\rangle$. From Fig.3, we know that two qubits is
separable if $\gamma{T}$ equals zero, which is different from the
case with the initial robust entangled state $|\Phi^{-}\rangle$.
However, the correspondence between the local maximal value of the
concurrence and the local minimal value of the information is
valid again. From Figs.1-3, we can observe that improving the
stationary state entanglement of two qubits implies the decrease
of the information and not vice versa. The correspondence between
the entanglement and information entropy of the stationary state
may help us to experimentally quantify the entanglement of two
qubits. In Ref.\cite{Ekert2002}, a scheme for direct estimations
of the functionals of the density matrix has been proposed. In
their scheme, one can extract certain properties such as the
Von-Neumann entropy of quantum states without recourse to quantum
tomography via a simple quantum network based on the
controlled-SWAP gate. Therefore, the correspondence between the
local maximal value of entanglement and local minimal value of
information entropy of the stationary state may provides us a more
efficient way to experimentally characterize the entanglement.

\section * {III. REMOTELY CONTROL THE DEPHASING OF TWO SOLID STATE QUBITS}

In this section, we investigate how to remotely control the
dephasing of two solid state qubits by making use of a preexistent
multi-partite entanglement and the quantum erasing process. We
assume that two solid state qubits are entangled with a remote
free qubit 3, such as polarized photon. Three qubits are initially
in the
$|GHZ\rangle=\frac{\sqrt{2}}{2}(|11\rangle_{1,2}|H\rangle_3+|00\rangle_{1,2}|V\rangle_3)$,
where $|H\rangle$ and $|V\rangle$ are two orthogonal states of the
qubit 3. If the qubit 3 is free out of any decoherence processes
and one of qubit 1 and qubit 2 in a collective dephasing channel
is driven by a finite-time external field, namely the evolution of
qubit 1 and qubit 2 are still governed by Eq.(3), the stationary
state of the three qubits can be obtained as follows: \beqa
\rho^{(s)}_{GHZ}&=&\zeta_a(\frac{\Omega_1}{\gamma},\gamma{T})|11\rangle_{1,2}\langle11|\otimes|H\rangle_3\langle{H}|\nonumber\\
&&+\zeta_b(\frac{\Omega_1}{\gamma},\gamma{T})|10\rangle_{1,2}\langle10|\otimes|V\rangle_3\langle{V}|\nonumber\\
&&+\zeta_c(\frac{\Omega_1}{\gamma},\gamma{T})|01\rangle_{1,2}\langle01|\otimes|H\rangle_3\langle{H}|\nonumber\\
&&+\zeta_d(\frac{\Omega_1}{\gamma},\gamma{T})|00\rangle_{1,2}\langle00|\otimes|V\rangle_3\langle{V}|\nonumber\\
&&+\zeta_f(\frac{\Omega_1}{\gamma},\gamma{T})|01\rangle_{1,2}\langle10|\otimes|H\rangle_3\langle{V}|\nonumber\\
&&+\zeta^{\ast}_f(\frac{\Omega_1}{\gamma},\gamma{T})|10\rangle_{1,2}\langle01|\otimes|V\rangle_3\langle{H}|,
\eeqa where the coefficients $\zeta_i$ ($i=a,b,c,d,f$) can be
calculated numerically according to Eq.(3). If the free qubit 3 is
simply traced out, the remaining two solid state qubits 1 and 2
have not any entanglement in the stationary state. In order to
maintain the entanglement of qubit 1 and qubit 2, we apply the
quantum erasing process to this system. The term "quantum eraser"
\cite{Scully1982} was invented to describe the loss or gain of
interference or, more generally quantum information, in a
subensemble, based on the measurement outcomes of two
complementary observables. It was reported that the implementation
of two- and three-spin quantum eraser using nuclear magnetic
resonance, and shown that quantum erasers provide a means of
manipulating quantum entanglement \cite{Rubin2000}. Recently, we
have shown how to entangle two mode thermal fields by utilizing
the quantum erasing \cite{Li2005}. The quantum erasing process
discussed here is implemented by measuring the polarizing vector
of the qubit 3. The project measurement of a polarized photon has
been extensively studied both in the theoretical and experimental
aspects. By making a project measurement of the qubit 3 on the
basis
$\{\cos\frac{\theta}{2}|H\rangle+e^{i\phi}\sin\frac{\theta}{2}|V\rangle,\cos\frac{\theta}{2}|V\rangle-e^{-i\phi}\sin\frac{\theta}{2}|H\rangle\}$,
the average pairwise entanglement characterized by concurrence in
the reduced density matrix of qubit 1 and qubit 2 can be obtained
\be
C_{ave}=2|\sin\theta|\max(0,|\zeta_f|-\sqrt{\zeta_a\zeta_d}).\ee
In Fig.4, we plot the average pairwise entanglement as the
function of $\gamma{T}$ and $\theta$. It is shown that an
appropriate value of the parameters $\gamma{T}$ and $\theta$ can
make two solid state qubits in a highly entangled state. One can
also change the parameters $\gamma{T}$ and $\theta$ to control the
entanglement of two qubits. Since the pairwise entanglement can
not arise if the qubit 3 is simply traced out, we can say that the
quantum erasing process realizes the remotely control of
entanglement of two solid state qubits. This may have some
potential applications in remotely control of decoherence.

\begin{figure}
\centerline{\includegraphics[width=2.5in]{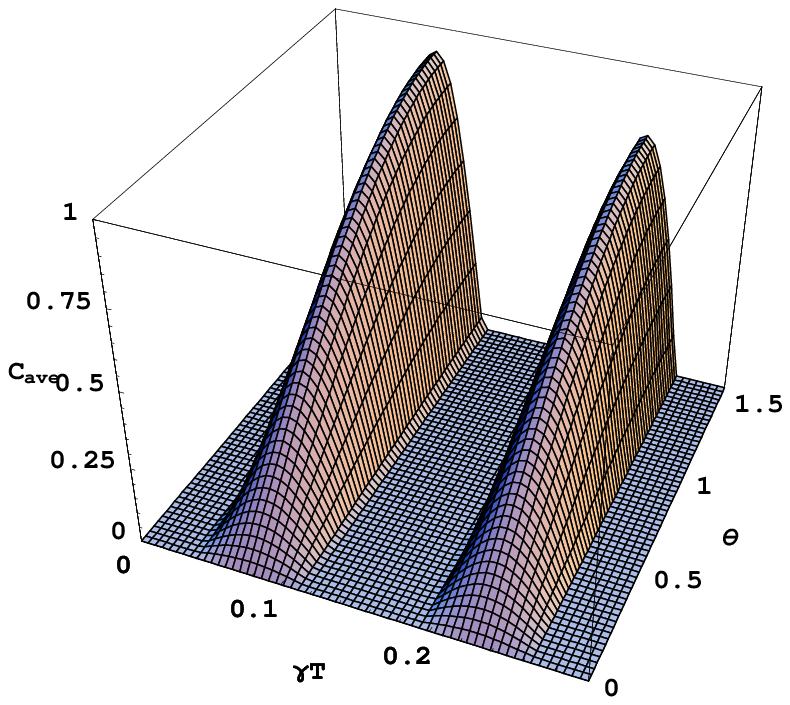}}
\caption{The average concurrence $C_{ave}$ of the stationary state
of qubit 1 and qubit 2 after the projection measurement on qubit 3
are plotted as the function of the parameter $\gamma{T}$ and
$\theta$ with $\frac{\Omega_1}{\gamma}=41.25$.}
\end{figure}

\section * {IV. CONCLUSIONS}

In this paper, we comparatively investigate the information
characterized by Von-Neumann entropy and the stationary state
entanglement quantified by concurrence of two locally driven solid
state qubits in the collective dephasing channel. We show that one
can transform the fragile entangled states into the stationary
entangled states under the collective dephasing by making use of a
finite-time external driving field. The local maximal value of the
stationary state concurrence corresponds to the local minimal
value of information. We show how a finite-time external driving
field can control the entanglement and information of the
stationary state of two qubits under the collective dephasing
environment. We also discuss the case that two qubits are
initially in the standard Werner state. It is shown that the
initial mixedness can change the stationary state entanglement and
information even though the other parameters of the model are
fixed. We also propose a scheme for remotely control the
entanglement of two solid state qubits via multi-partite
entanglement and the quantum erasing process. This idea may shed
some light on the universal quantum operation based on
decoherence-free qubits, and provide us a possible way to remotely
suppress the detrimental effects of decoherence. Our results may
have potential applications in quantum teleportation
\cite{Bennett1993,Bouwmester1997} or other remote quantum
information processes. In the future work, it may be very
interesting to apply the present results to
some realistic quantum information processes.\\

\section*{ACKNOWLEDGMENT}
This project was supported by the National Natural Science
Foundation of China (Project NO. 10174066).

\bibliographystyle{apsrev}
\bibliography{refmich,refjono}

\end{document}